\documentclass[%
reprint,
 amsmath,amssymb,
 aps,
]{revtex4-1}

\usepackage{graphicx}
\usepackage{dcolumn}
\usepackage{amsmath}
\usepackage{amssymb}
\usepackage{bm}
\usepackage{subfigure}
\usepackage{slashed}
\usepackage{hyperref}
\usepackage{mathrsfs}

\usepackage{tikz}
\usetikzlibrary{shapes,shadows,arrows,positioning,calc,intersections}
\tikzstyle{block}=[draw,rectangle, text centered, node distance=7em]
\tikzstyle{line}=[draw,-latex]

\usepackage{bm}

\def\p{\mbox{\boldmath$\displaystyle\mathbf{p}$}}

\def\0{\mbox{\boldmath$\displaystyle\mathbf{0}$}}

\begin{document}


\title{Spin and helicity of massive particles}

\author{Cheng-Yang Lee}
\email{chengyang.lee45@gmail.com}
\affiliation{
Institute of Mathematics, Statistics and Scientific Computation,\\
Unicamp, 13083-859 Campinas, S\~{a}o Paulo, Brazil 
}%
\affiliation{
Department of Physics and Astronomy,\\
Rutherford Building, University of Canterbury \\ 
Private Bag 4800, Christchurch 8140, New Zealand
}

\date{\today}

\begin{abstract}
It is a well-known fact that helicity is a Lorentz-invariant for massless but not for massive particles. Nevertheless, a satisfactory proof of this fact and a detailed analysis on the relative orientation between spin and the momentum are not readily available. One such analysis was presented by Wigner where a formula for the angle between the spin and the momentum is derived for massive particle states undergoing two successive boosts in orthogonal directions. In this note, using the Wigner rotation, an alternative derivation is provided.
\end{abstract}

\pacs{12.60.Fr}
\maketitle


\section{Introduction}

A well-known fact in quantum field theory is that the number of allowed degenerate states for massless and massive particles belonging to a spin-$j$ representation of the Poincar\'{e} group are labelled by the spin projection $\sigma=\pm j$ and $\sigma=-j,\cdots,j$ where $j=0,\frac{1}{2},1,\cdots$~\cite{Wigner:1939cj}. For massless particles, they are helicity eigenstates in all reference frames where the spin is always parallel or anti-parallel to the momentum. As for massive particles, while it is possible to prepare helicity eigenstates, they are not Lorentz-invariant.

While these statements are well-known and are often repeated in textbooks, a detailed analysis on the relative orientation between the spin and the momentum is not readily available. To the best of our knowledge, such analysis was first presented by Wigner in ref.~\cite[Chap.~5]{Wigner:essay} along with more recent works presented in refs.~\cite{Polyzou:2012ut,Duval:2014cfa,Duval:2014ppa,Elbistan:2015bha}. We are particularly interested in Wigner's analysis where the notion of an angle between the spin and the momentum is introduced. As an example, Wigner considered a massive particle state that undergoes two successive boosts in orthogonal directions and obtained a formula for the angle.

In this note, we start with the simplest examples of single rotation and boost. Subsequently, we consider the case of successive boosts in orthogonal direction. We provide an alternative but equivalent derivation to Wigner's result (see app.~\ref{A}), utilizing the Wigner rotation.

\section{Massive states}

In the rest frame, we define the massive state $|k,\sigma\rangle$ to be an eigenstate of the rotation generator $J_{z}$ along the $z$-axis
\begin{equation}
J_{z}|k,\sigma\rangle=\sigma|k,\sigma\rangle \label{eq:spin}
\end{equation}
where $k^{\mu}=(m,\0)$ with $m$ being the mass of the particle. This equation cannot be Lorentz-invariant since $J_{z}$ is not a Casimir-invariant. Therefore, to understand the relative orientation between the spin and the momentum, we need to consider the particle states in various reference frames. The Lorentz transformation for a massive particle state is given by~\cite{Weinberg:1995mt}
\begin{equation}
U(\Lambda)|p,\sigma\rangle=\sqrt{\frac{(\Lambda p)^{0}}{p^{0}}}\sum_{\sigma'}D_{\sigma'\sigma}(W(\Lambda,p))|\Lambda p,\sigma'\rangle\label{eq:trans}
\end{equation}
where $U(\Lambda)$ is the unitary representation of $\Lambda$ and the normalization is chosen to be
\begin{equation}
\langle p',\sigma'|p,\sigma\rangle=\delta_{\sigma'\sigma}\delta^{3}(\p'-\p).
\end{equation}
In eq.~(\ref{eq:trans}), $W(\Lambda,p)=L^{-1}(\Lambda p)\Lambda L(p)$ is an element of the little group defined to leave $k^{\mu}$ invariant $Wk=k$, $L(p)$ is a boost that takes particle at rest to momentum $p$ and the matrix $D(W(\Lambda,p))$ is the unitary representation of the little group. For massive particle, the little group is $SU(2)$.

When $\Lambda=R$ is a rotation, we have $W(R,p)=R$. For a boost $\Lambda=L(q)$ where $p\neq q$, $W(L(q),p)$ remains a rotation but the expression is more complicated. A practical method which we will use in sec.~\ref{B} to compute $W(L(q),p)$ can be found in ref.~\cite{Ferraro:1999eu}. 

In the subsequent sections we study the relative orientation of the spin and the momentum for massive particle states when they undergo Lorentz transformations. The analysis for rotation and single boost are  straightforward, but for the sake of completeness, we will review them here. The analysis becomes more interesting for two successive boosts in orthogonal directions.

\subsection{Rotation}

The rotation on the state $|k,\sigma\rangle$ is given by
\begin{equation}
U(R)|k,\sigma\rangle=\sum_{\sigma'}D_{\sigma'\sigma}(R)|k,\sigma'\rangle.
\end{equation}
The rotated state is an eigenstate of $U(R)J_{z}U^{-1}(R)=\boldsymbol{J\cdot n}$
 so we get
\begin{equation}
\boldsymbol{J\cdot n}\,U(R)|k,\sigma\rangle=\sigma U(R)|k,\sigma\rangle
\end{equation}
where $\mathbf{n}$ specifies the direction of spin. 

\subsection{Boost} \label{B}
Transformations under  boosts are more complicated since they do not commute with rotation. From eq.~(\ref{eq:trans}), a boost on $|k,\sigma\rangle$ yields
\begin{equation}
|p,\sigma\rangle=\sqrt{\frac{m}{p^{0}}}U(L(p))|k,\sigma\rangle.\label{eq:psigma}
\end{equation}
When $\p\neq (0,0,p_{z})$, $|p,\sigma\rangle$ is not a helicity eigenstate of $J_{z}$ even though the label $\sigma$ remains unchanged in eq.~(\ref{eq:psigma}). According to Wigner, there is a non-vanishing angle $\epsilon$ between the spin and the momentum. Qualitatively, this means that prior to applying the boost, one can perform a rotation by  angle $\epsilon$ such that the final boosted state $U(L(p)R(\epsilon))|k,\sigma\rangle$ is a helicity eigenstate. In  this paper, we derive the formulae for $\epsilon$ in the case of a single boost and subsequently two successive boosts in orthogonal directions.

We start with the simplest case where a particle is moving along the $z$-axis with momentum $\p=(0,0,p_{z})$. This is a helicity eigenstate whose spin is parallel to its momentum so we have
\begin{equation}
\mathbf{J}\cdot\hat{\p}|p_{z},\sigma\rangle=\sigma|p_{z},\sigma\rangle.
\end{equation}
It is straightforward to see that the helicity is not Lorentz-invariant by going to another frame where the particle has momentum $\p'=(0,0,-p'_{z})$ so that the spin becomes anti-parallel to the momentum
\begin{equation}
\mathbf{J\cdot\hat{\p}'}|p'_{z},\sigma\rangle=-\sigma|p'_{z},\sigma\rangle.
\end{equation}

Now, what happens if we apply a single boost to $|k,\sigma\rangle$ in the $yz$-plane $L(\theta,\varphi)$, where the momentum makes an angle $\theta$ with the $z$-axis? To answer this question, we use the identity
\begin{equation}
R(\theta)L(0,\varphi)R^{-1}(\theta)=L(\theta,\varphi)\label{eq:r}
\end{equation}
where $R(\theta)$ is the rotation about the $x$-axis
\begin{equation}
R(\theta)=\left(\begin{matrix}
1 & 0 & 0 & 0 \\
0 & 1 & 0 & 0 \\
0 & 0 & \cos\theta & \sin\theta \\
0 & 0 & -\sin\theta & \cos\theta
\end{matrix}\right)
\end{equation} 
and $L(0,\varphi)$ is a boost along the $z$-axis
\begin{equation}
L(0,\varphi)=\left(\begin{matrix}
\cosh\varphi & 0 & 0 & \sinh\varphi \\
0 & 1 & 0 & 0 \\
0 & 0 & 1 & 0 \\
\sinh\varphi & 0 & 0 & \cosh\varphi
\end{matrix}\right).\label{eq:ltp}
\end{equation}
Starting from
\begin{equation}
J_{z}U(L(0,\varphi))|k,\sigma\rangle=\sigma U(L(0,\varphi))|k,\sigma\rangle,
\end{equation}
we get
\begin{equation}
J_{yz} U(L(\theta,\varphi))U(R)|k,\sigma\rangle=\sigma U(L(\theta,\varphi))U(R)|k,\sigma\rangle
\end{equation}
where we have used $U(R)J_{z}U^{-1}(R)=J_{yz}$ with $J_{yz}$ being the rotation generator in the $yz$-plane. 
Note that although $U(L(\theta,\varphi))|k,\sigma\rangle$ is not a helicity eigenstate but $U(L(\theta,\varphi))U(R)|k,\sigma\rangle$ is. Therefore, in the case of a single boost, the angle between the spin and the momentum is $\epsilon=\theta$ and it can be expressed in terms of the components of the velocity along the $y$- and $z$-axis
\begin{equation}
\tan\epsilon=\tan\theta=\frac{\beta_{y}}{\beta_{z}}\label{eq:sb}
\end{equation}
where $\p=m\gamma_{yz}(0,\beta_{y},\beta_{z})$. For example, if the boost is along the $y$-axis, we have $\beta_{z}=0$ and $\epsilon=\pi/2$. 
Since there are no preferred frames, it follows that for a massive particle state at rest where its spin aligned to an arbitrary direction, its orientation remains unchanged under a single boost in any direction.

\begin{widetext}
Now let us consider a state that is obtained by acting two successive boosts on $|k,\sigma\rangle$, first along the $z$- and then along the $y$-axis. The product of the two boosts are
\begin{eqnarray}
L(\textstyle{\frac{1}{2}\pi},\varphi')L(0,\varphi)&=&
\left(\begin{matrix}
\cosh\varphi' & 0 & \sinh\varphi' & 0 \\
0 & 1 & 0 & 0 \\
\sinh\varphi' & 0 & \cosh\varphi' & 0 \\
0 & 0 & 0 & 1
\end{matrix}\right)
\left(\begin{matrix}
\cosh\varphi & 0 & 0 & \sinh\varphi \\
0 & 1 & 0 & 0 \\
0 & 0 & 1 & 0 \\
\sinh\varphi & 0 & 0 & \cosh\varphi
\end{matrix}\right)\nonumber\\
&=&\left(\begin{matrix}
\cosh\varphi'\cosh\varphi & 0 & \sinh\varphi' & \cosh\varphi'\sinh\varphi \\
0 & 1 & 0 & 0 \\
\sinh\varphi'\cosh\varphi & 0 & \cosh\varphi' & \sinh\varphi'\sinh\varphi \\
\sinh\varphi & 0 & 0 & \cosh\varphi\label{eq:two_boosts}
\end{matrix}\right).
\end{eqnarray}
To obtain the angle between the spin and the momentum, we use the fact that two boosts along different directions is equivalent to a boost and a rotation. Specifically, eq.~(\ref{eq:two_boosts}) can be written as
\begin{equation}
L(\textstyle{\frac{1}{2}\pi},\varphi')L(0,\varphi)=L(\theta'',\varphi'')R(\theta_{w})
\end{equation}
where $L(\theta'',\varphi'')$ is a boost along the $yz$-plane and $R(\theta_{w})$ is the element of the little group. To determine $L(\theta'',\varphi'')$, we note that $L(\textstyle{\frac{1}{2}\pi},\varphi')L(0,\varphi)$ leaves the $x$-axis invariant. Therefore, $R(\theta_{w})$ must be a rotation about the $x$-axis
so we get
\begin{equation}
L(\theta'',\varphi'')=\left(\begin{matrix}
\cosh\varphi'\cosh\varphi & 0 & \cos\theta_{w}\sinh\varphi'+\cosh\varphi'\sin\theta_{w}\sinh\varphi &
\cos\theta_{w}\cosh\varphi'\sinh\varphi-\sin\theta_{w}\sinh\varphi' \\
0 & 1 & 0 & 0 \\
\cosh\varphi\sinh\varphi' & 0 & \cos\theta_{w}\cosh\varphi'+\sin\theta_{w}\sinh\varphi'\sinh\varphi &
\cos\theta_{w}\sinh\varphi'\sinh\varphi-\sin\theta_{w}\cosh\varphi' \\
\sinh\varphi & 0 & \cosh\varphi \sin\theta_{w} & \cos\theta_{w}\cosh\varphi
\end{matrix}\right).\label{eq:varphi2}
\end{equation}
\end{widetext}
For $L(\theta'',\varphi'')$ to qualify as a boost, the matrix must be symmetric. Equating all the relevant elements, we obtain  \begin{eqnarray}
&&\sin\theta_{w}=\frac{\sinh\varphi'\sinh\varphi}{\cosh\varphi'\cosh\varphi+1}, \\
&&\cos\theta_{w}=\frac{\cosh\varphi'+\cosh\varphi}{\cosh\varphi'\cosh\varphi+1}
\end{eqnarray}
so that
\begin{eqnarray}
\tan\theta_{w}&=&\frac{\sinh\varphi'\sinh\varphi}{\cosh\varphi'+\cosh\varphi}\nonumber\\
&=&\frac{\gamma\gamma'\beta\beta'}{\gamma+\gamma'}.
\end{eqnarray}
When $U(L(\theta'',\varphi'')R(\theta_{w}))$ acts on the state $|k,\sigma\rangle$, after rotation, the spin makes an angle $\theta_{w}$ with the $z$-axis. Therefore, if $\theta''=\theta_{w}$, $U(L(\theta'',\varphi'')R(\theta_{w}))|k,\sigma\rangle$ would be a helicity eigenstate. But in general, the two angles do not coincide so the angle between the spin and the momentum is $\epsilon=\theta''-\theta_{w}$. From eq.~(\ref{eq:varphi2}), the $\gamma$-factor and the velocity components are
\begin{eqnarray}
&& \gamma''=\gamma\gamma',\\
&& \beta''_{y}=\beta',\\
&& \beta''_{z}=\frac{\beta}{\gamma'}.
\end{eqnarray}
Using these identifications, $\theta''$ is given by
\begin{equation}
\tan\theta''=\frac{\beta''_{y}}{\beta''_{z}}=\gamma'\frac{\beta'}{\beta}
\end{equation}
and the angle between the spin and the momentum after boost in the $z$- and subsequently the $y$-axis is given by
\begin{equation}
\tan\epsilon=\frac{\tan\theta''-\tan\theta_{w}}{1+\tan\theta''\tan\theta_{w}}.\label{eq:ep}
\end{equation}
At first sight, eq.~(\ref{eq:ep}) looks different from eq.~(\ref{eq:angle}) but after some algebraic manipulation, they are identical
\begin{equation}
\tan\epsilon=\frac{\tanh\varphi'}{\sinh\varphi}=\frac{\beta'}{\beta}(1-\beta^{2})^{1/2}.
\end{equation}
When $\beta=1$, we have a massless particle moving along the $z$-axis. Independent of the subsequent boost, $\epsilon=0$. Therefore, we have a massless helicity eigenstate where its spin is always parallel or anti-parallel to the momentum. In the non-relativistic limit, $\theta_{w}\sim 0$ so we get $\tan\epsilon\sim\beta'/\beta$.  

The prescription we have used to derive the angle between the spin and the momentum from the little group can be generalized to various situations. The important insight is that successive boosts in different directions can be decomposed into product of boost and rotation and that the angle can be derived by determining the orientations of the two operations.

\section*{Acknowledgements}

This research is supported by the CNPq grant 313285/2013-6. I would like to thank the generous hospitality offered by the Department of Physics and Astronomy of the University of Canterbury where part of this work was completed.

\appendix

\section{Wigner's derivation} \label{A}
Wigner's derivation for the angle between the spin and the momentum under two successive boosts along the $z$- and then the $y$-axis can be found in ref.~\cite[Chap.~5]{Wigner:essay}. For the reader's convenience, it is reproduced here.

First, we prepare a helicity eigenstate moving in the $yz$-plane making an angle $\theta$ with the $z$-axis.
The state can be prepared by first performing a rotation about the $x$-axis followed by a boost in the same direction which is given by eq.~(\ref{eq:r}). This transformation takes the form
\begin{eqnarray}
\hspace{-0.5cm}T(\theta,\varphi)&=&L(\theta,\varphi)R(\theta)\nonumber\\
&=&\left(\begin{matrix}
\cosh\varphi & 0 & 0 & \sinh\varphi \\
0 & 1 & 0 & 0 \\
\sin\theta \sinh\varphi & 0 & \cos\theta & \sin\theta\cosh\varphi \\
\cos\theta\sinh\varphi & 0 & -\sin\theta & \cos\theta \cosh\varphi
\end{matrix}\right).\label{eq:T}
\end{eqnarray}

\begin{widetext}
Wigner asserts that the state $U(L(\frac{1}{2}\pi,\varphi')L(0,\varphi)R(\epsilon))|k,\sigma\rangle$ is a helicity eigenstate because it can be equated to $U(T(\theta,\varphi''))|k,\sigma\rangle$ for some rapidity parameter $\varphi''$. Therefore, the angle between the spin and the momentum is $\epsilon$. To determine $\epsilon$, we multiply eq.~(\ref{eq:two_boosts}) from the right by $R(\epsilon)$ to obtain
\begin{eqnarray}
L(\textstyle{\frac{1}{2}\pi},\varphi')L(0,\varphi)R(\epsilon)=\left(\begin{matrix}
\cosh\varphi'\cosh\varphi & 0 & (\cos\epsilon\sinh\varphi'-\sin\epsilon\cosh\phi'\sinh\varphi) &
(\sin\epsilon\sinh\varphi'+\cos\epsilon\cosh\varphi'\sinh\varphi) \\
0 & 1 & 0 & 0\\
\cosh\varphi\sinh\varphi' & 0 & (\cos\epsilon\cosh\varphi'-\sin\epsilon\sinh\varphi'\sinh\varphi) &
(\sin\epsilon\cosh\varphi'+\cos\epsilon\sinh\varphi'\sinh\varphi) \\
\sinh\varphi & 0 &-\sin\epsilon\cosh\varphi & \cos\epsilon\cosh\varphi
\end{matrix}\right).\nonumber\\
\label{eq:LL}
\end{eqnarray}
\end{widetext}
Equation~(\ref{eq:LL}) then take the form of (\ref{eq:T}) when
\begin{equation}
\tan\epsilon=\frac{\tanh\varphi'}{\sinh\varphi}=\frac{\beta'}{\beta}(1-\beta^{2})^{1/2} \label{eq:angle}
\end{equation}
so that $\epsilon$ is the angle between the spin and the momentum of the state. When $\beta=1$, $\epsilon=0$ so we have a massless helicity eigenstate. In the non-relativistic limit,  $\tan\epsilon\sim\beta'/\beta$.

\bibliography{Bibliography}

\begin{thebibliography}{1}

\bibitem{Wigner:1939cj}
Eugene~P. Wigner.
\newblock {On unitary representations of the inhomogeneous Lorentz group}.
\newblock {\em Annals Math.}, 40:149--204, 1939.

\bibitem{Wigner:essay}
Eugene~P. Wigner.
\newblock {Symmetries and reflections: Scientific essays of Eugene P. Wigner}.
\newblock {\em Indiana Univ. Pr.}, page 280 p, 1967.

\bibitem{Polyzou:2012ut}
W.~N. Polyzou, W.~Glöckle, and H.~Witala.
\newblock {Spin in relativistic quantum theory}.
\newblock {\em Few Body Syst.}, 54:1667--1704, 2013.

\bibitem{Duval:2014cfa}
C.~Duval, M.~Elbistan, P.~A. Horváthy, and P.~M. Zhang.
\newblock {Wigner–Souriau translations and Lorentz symmetry of chiral
  fermions}.
\newblock {\em Phys. Lett.}, B742:322--326, 2015.

\bibitem{Duval:2014ppa}
C.~Duval and P.~A. Horvathy.
\newblock {Chiral fermions as classical massless spinning particles}.
\newblock {\em Phys. Rev.}, D91(4):045013, 2015.

\bibitem{Elbistan:2015bha}
M.~Elbistan, C.~Duval, P.~A. Horvathy, and P.~M. Zhang.
\newblock {Helicity of spin-extended chiral particles}.
\newblock {\em Phys. Lett.}, A380:1677--1683, 2016.

\bibitem{Weinberg:1995mt}
Steven Weinberg.
\newblock {The quantum theory of fields. Vol. 1: Foundations}.
\newblock {\em Cambridge, UK: Univ. Pr.}, page 609 p, 1995.

\bibitem{Ferraro:1999eu}
R.~Ferraro and M.~Thibeault.
\newblock {Generic composition of boosts: an elementary derivation of the
  Wigner rotation}.
\newblock {\em Eur. J. Phys.}, 20:143--151, 1999.

\end{thebibliography}
 \bibliographystyle{unsrt}


\end{document}